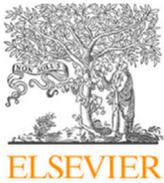
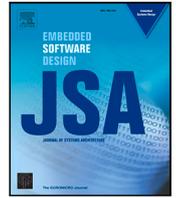

# SLIMBRAIN: Augmented reality real-time acquisition and processing system for hyperspectral classification mapping with depth information for in-vivo surgical procedures

Jaime Sancho [a],[*], Manuel Villa [a], Miguel Chavarrías [a], Eduardo Juarez [a], Alfonso Lagares [b], César Sanz [a]

[a] *Research Center on Software Technologies and Multimedia Systems, Universidad Politécnica de Madrid, Spain*
[b] *Department of Neurosurgery, Hospital Universitario 12 de Octubre, Universidad Complutense de Madrid, Spain*

## A R T I C L E   I N F O

## A B S T R A C T

Over the last two decades, augmented reality (AR) has led to the rapid development of new interfaces in various fields of social and technological application domains. One such domain is medicine, and to a higher extent surgery, where these visualization techniques help to improve the effectiveness of preoperative and intraoperative procedures. Following this trend, this paper presents SLIMBRAIN, a real-time acquisition and processing AR system suitable to classify and display brain tumor tissue from hyperspectral (HS) information. This system captures and processes HS images at 14 frames per second (FPS) during the course of a tumor resection operation to detect and delimit cancer tissue at the same time the neurosurgeon operates. The result is represented in an AR visualization where the classification results are overlapped with the RGB point cloud captured by a LiDAR camera. This representation allows natural navigation of the scene at the same time it is captured and processed, improving the visualization and hence effectiveness of the HS technology to delimit tumors. The whole system has been verified in real brain tumor resection operations.

## 1. Introduction

Over the last two decades, augmented reality (AR) has led to the rapid development of new interfaces in various fields of social and technological application domains [1–4]. The motivations for this development range from the entertainment industry to the remote control of industrial installations. In the field of medicine, the use of AR is employed to aid medical personnel before or during operations by providing more accurate diagnoses. However, especially in the healthcare field, AR alone may not be worthwhile as a contribution in itself [5]. Nonetheless, this work focuses on contributing to the combined use of advanced imaging techniques in the healthcare domain, data analysis, and real-time information processing.

Therefore, the possibility of providing rapid and minimally invasive diagnostic systems is of paramount importance. In this respect, the use of hyperspectral (HS) imaging (HSI) has proved its worth in characterizing, employing spectrography, the different types of tissues captured in an image. The use of this technology has not been limited to medicine. Additionally, it has been used in fields such as remote sensing, mining or agriculture [6–8]. The challenge lies in the precise characterization of human tissues to specific spectral signatures. The enormous natural variability of such substances makes direct classification by spectrography unfeasible. For this reason, the information from the HS cameras is used to build models based on machine learning techniques capable of extracting and detecting patterns that, henceforth, yield valuable results [9].

As a counterpart, these techniques entail huge computational loads, which makes difficult their implementation in real-time. Many efforts have been driven in this regard, using hardware accelerators such as Graphics Processing Units (GPU) to obtain results that can be shown during the course of a real operation [10]. However, AR applications demand not only a classification image to present but an interactive video streaming that can produce an immersive experience. This fact leads to the need of generating HS classification maps in video real-time as well as the so-called depth maps, needed to generate the AR scene.

This paper presents SLIMBRAIN, a real-time acquisition, and processing system for HS classification with depth information. This system has been developed using as a starting point HS imaging in medicine for the delimitation of brain tumors. Thanks to the combined use of HS and machine learning techniques, it is possible to generate classification maps that identify the tissue typology of the patient undergoing surgery. However, this information lacks both volumetric information and realism. In this work, the immersive information comes from a LiDAR device equipped with a laser and an RGB camera. The combined use of LiDAR and HS imaging has been used before in remote sensing. However, to the authors' knowledge, this is the first time it has been used in the field of image-assisted surgery. In addition, to support a continuous classification tissue flow, a snapshot HS camera has been used. A fundamental part of the work consisted of the development of a toolchain that encompasses the processing of the HS images, the generation of the classification maps, the registration of these maps with the depth information, and the continuous rendering of a video stream in an immersive way. Finally, the framework has been implemented in such a way that it is capable of providing real-time response on a standard high-performance gaming workstation, mainly relying on GPU processing.

The rest of the paper is organized as follows: In Section 2 the background of this work is presented, focusing on the HS imaging use in AR. Section 3 details the proposed solution with emphasis on both, the acquisition system and the image processing chain. The verification of the system is presented in Section 4; it has been carried out first in a laboratory and then in a real application environment, i.e. a neurosurgical operating theater, using a high-performance processing platform. Finally, Section 5 draws the conclusions of this work.

**2. State-of-the-art**

This work lays its foundations in three main fields: (i) HS and LiDAR data fusion, (ii) real-time HSI for medical applications, and (iii) AR in the medical field. The current state-of-the-art (SoTA) for them is analyzed in the following subsections.

*2.1. Hyperspectral and LiDAR data fusion*

HS and LiDAR data fusion has been widely explored in the state-of-the-art (SoTA) [11] mainly in the domain of remote sensing. However, that application determines many technical aspects that differentiate this work from the SoTA. One aspect is how the LiDAR information is utilized. It is employed to improve to some extent the classification by adding geometric information to the HS image. In this way, authors are able to identify buildings [12] or man-made objects [13], among others. In this work, the LiDAR depth information is not employed as a feature to improve the classification but as a way to improve the visualization of the HS classification by generating an AR system.

Another aspect is the HS cameras and LiDARs employed. For remote sensing the vast majority of HS cameras employed are pushbroom and the LiDARs utilized are long-range. In this work, the application requires a real-time capture of the scene so the AR system can be interactive and immersive. For this reason, the HS camera employed is a snapshot and the LiDAR is a close-ranged camera able to produce video streaming at a rate of up to 30 frames per second (FPS).

The last aspect is the real-time constraint. Whilst in remote sensing the image classification and processing can be done offline, this application requires real-time processing as much close as possible to the capture rate. For this reason, unlike in remote sensing, for this work, the GPU acceleration of the process is considered a key factor.

*2.2. Real-time HSI for medical applications*

There are many works successfully addressing the use of HSI for detecting diseases such as breast [14], lung [15] or brain cancer [9,16]. However, it is difficult to find works in the state-of-the-art focusing on a real-time video implementation of them mainly because the cameras employed in that works are linescan type. These cameras are not able to capture a video stream, as they need to be moved over a scene to obtain a single capture. These captures feature high spatial and spectral resolution, with the inconvenience of a capture time of up to minutes. During the last years, commercial HS snapshot cameras [17] appeared as a solution to the capture time problem, with the counterpart of a lower spatial and spectral resolution.

HS snapshot cameras manufacturers are clearly targeting video real-time applications [18] with products with up to 170 raw HS cubes per second [19]. These cameras count with a reduced number of spectral bands, compared to HS linescan-type cameras and hence, their range of application is more limited. However, they still can provide useful information about the materials present in the scene, depending on the case of study [20]. One example is the work of Van Manen et al. [21], where a snapshot camera that acquires 41 spectral bands ranging from 470 nm to 950 nm is employed for detecting cutaneous oxygen saturation. Their results prove the feasibility of the system for that purpose. Another example was presented by Kaluzny et al. [22], with a camera featuring 16 spectral bands between 460 nm and 630 nm for retinal oximetry and macular pigment optical density measurements. They proved the usefulness of snapshot cameras in ophthalmology, emphasizing the advantages of capturing instantaneous HS cubes. As a final example, He and Wang presented in their work [23] a novel analysis of skin morphological features that benefits from using a real-time HSI system. In their work, they are able to monitor blood or melanocytic nevus absorption levels using the properties of a HS camera. However, thanks to the HS real-time acquisition, they can also measure the heart rate, the exercise recovery rate, or the vascular occlusion.

These works demonstrate the ability of HS snapshot cameras to capture valuable information in the medical ambit, even if the number of captured bands is reduced. The capacity of capturing in video real-time opens a new research dimension, offering the possibility of a diagnostic based on a temporal variation or easing the current procedure in solutions based on HS cameras.

*2.3. AR and multimodal image systems in the medical field*

Nowadays, AR is a technology with an increasing impact in many research areas such as education [24], manufacturing [25] o training [26]. In the context of medical imaging, only in the year 2020, 802 works were published, presenting an exponential increment compared with the previous years [5]. These numbers are reflected in papers such as [27], showing many examples of successful solutions in the domain of neurosurgery. These solutions mainly improve surgeons' criteria by adding valuable information about the tissues inspected or by reducing the preoperative planning time. For example, in the context of brain tumor resection, several works [28,29] employ the information from a preoperative Magnetic Resonance (MR) to be displayed over the exposed brain. In this way, the distinguishable tumor is presented in order to help neurosurgeons to locate and resect it. Another example is the work in [30], where a preoperative computed tomography model is employed as an intraoperative guidance method to improve orthopedic surgical operations. These work, among others, [31–35], follows the direction of displaying an AR where preoperative information captured by other sources and sometimes processed, enhances the reality observed by the medical personnel. Normally, this information from the inside of the human body is represented on its surface, enabling better preparation or easing the surgery. This shows a popular trend where medical AR could play an important role in the following years.

Although the proposal in this work has a similar approach to the aforementioned solutions, it extends the challenges to solve by displaying an AR captured and processed intraoperatively. Taking into account this aspect, the state-of-the-art is not as populated as before. In [36], authors present a system that uses fluorescence lifetime imaging in real-time to generate an AR. With this tool, they are able to provide diagnostic information on the image captured by the surgical microscope without perturbing the normal course of the operation. Similarly, in [37], the image obtained by a 2D endoscopic image is enhanced and presented to the doctors. In this case, the image is segmented using a CNN and a 3D model is displayed over the 2D image in order to guide the surgeons during radical prostatectomy operations. Another example is [38], where a multimodal AR in a digital surgical microscope is proposed. Their proposal employs a stereoscopic multispectral imaging system that enables both the 3D reconstruction and tissue classification of the patient region of interest. With that information, they are able to generate an AR that helps to detect and delimit different types of tissues.

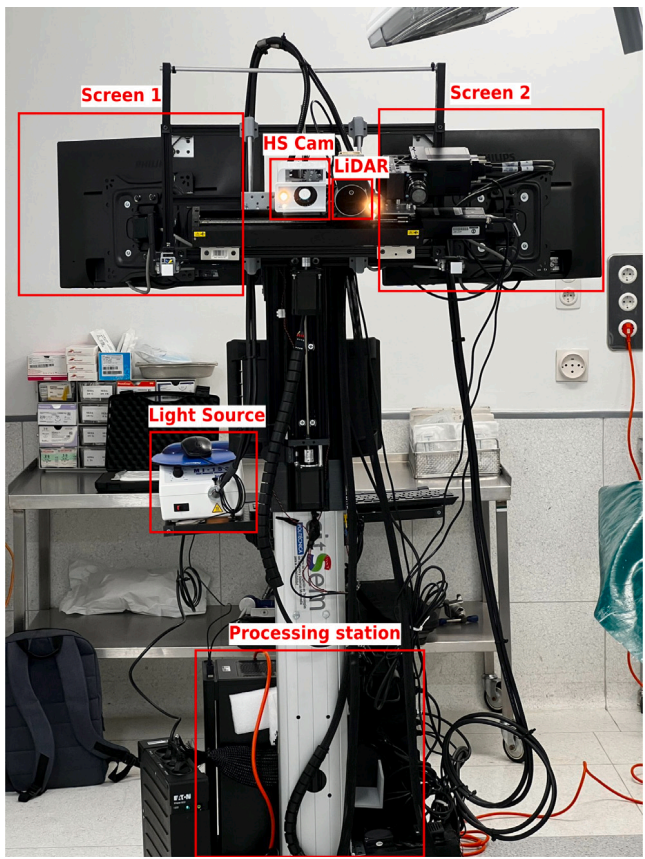

**Fig. 1.** Slimbrain physical system. The main elements are a HS camera, a LiDAR, a light source, a processing station, and two screens.

These works present many research efforts aiming to improve both preoperative and intraoperative imaging techniques by developing multimodal systems where the information is presented in AR. This initiative still has technical challenges to overcome, however, it seems clear that there exists a necessity of improving immersive medical imaging systems.

## 3. Slimbrain

Slimbrain is a real-time acquisition and processing system for HS classification employing AR. This system presents to the neurosurgeons a freely-navigable 3D point cloud video captured in situ where the tumor tissue is highlighted and differentiated from the rest of the tissues of the patients. The physical system, presented in Fig. 1, is further explained in Section 3.1. The full processing chain, depicted in Fig. 2, is divided into three main stages: (i) HS processing chain in Section 3.2, (ii) depth processing chain in Section 3.3 and (iii) registration and rendering in Section 3.4.

### 3.1. Setup: hardware and devices

In order to fulfill the application requirements, the physical system needs to count with several devices. They are listed below along with the requirement that they fulfill:

1. A HS snapshot camera, Ximea MQ022HG-IM-SM5X5-NIR2, with 25 spectral bands ranging from 665 nm to 960 nm and a resolution of 409 × 217 spectral pixels. This camera presents a 2045 × 1085 sensor able to acquire up to 170 raw FPS plus a spectral 5 × 5 mosaic filter in front of it. In this way, a 409 × 217 × 25 HS cube can be generated by demosaicing the filtered bands. This HS camera suits the application requirements as it is able to produce valuable HS information at real-time video frame rates.

2. A special light source, Dolan-Jenner Model Mi-150, able to illuminate all the wavelengths captured by the previous HS camera.

3. An Intel Realsense L515 LiDAR composed of a indirect Time-of-Flight (ToF) camera and an RGB camera. The depth sensor has an average depth accuracy between 5 mm and 14 mm and a resolution of 1024 × 768 resolution pixels, with a frame rate of up to 30 FPS. The RGB camera sensor has full HD resolution and also works up to 30 FPS. It is important to remark that the RGB and the depth sensor are different cameras and in different positions (around 14 mm apart); this means that the images captured are not aligned and cannot be directly employed as an RGB-depth (RGBD) capture, i.e., a color image with depth information for each pixel. This LiDAR provides RGB and depth information at the real-time frame rate required by the application in a close range (between 0.25 m and 1 m with a depth accuracy of 5 mm).

4. A processing station consisting of an Intel core i9 10900K CPU plus an NVIDIA RTX3090 GPU. In this work, the GPU allows for successfully exploiting the inherent parallelism in image pixels and hence is of paramount importance for the acceleration. The CPU and GPU selection pursues the achievement of real-time processing of the information captured.

5. Two screens that allow the control of the system and the display of the output.

The scheme included in Fig. 3 depicts the data flow from the aforementioned sources through the processing station. As can be seen in the figure, the information flows from the left to the right, starting from the three cameras in the system. For each frame, all the images are synchronously captured and stored in the CPU memory. Then, they are transferred without employing any tiling or batching technique to the GPU (processing time results without using these techniques are enough for this application), where they are processed (this is explained in Sections 3.2, 3.3, and 3.4) to generate a point-cloud representation from them. Finally, the frame is directly rendered on the screens through the video output of the GPU.

The selection of these elements to conform the final system has been performed through a process of comparing the hardware specifications of the different options available on the market. However, no other hardware elements have been tested in the system as the ones chosen suits the application requirements.

### 3.2. HS processing chain

The HS processing chain is based on previous works [39,40], with several improvements. These works took special attention to the data arrangement and the data type in order to enable the use of the HS processing chain in embedded devices and eased the inclusion of new stages. The new processing chain is composed of four steps: (i) pre-processing, (ii) supervised classification, (iii) unsupervised classification, and (iv) majority voting (upper part of Fig. 2).

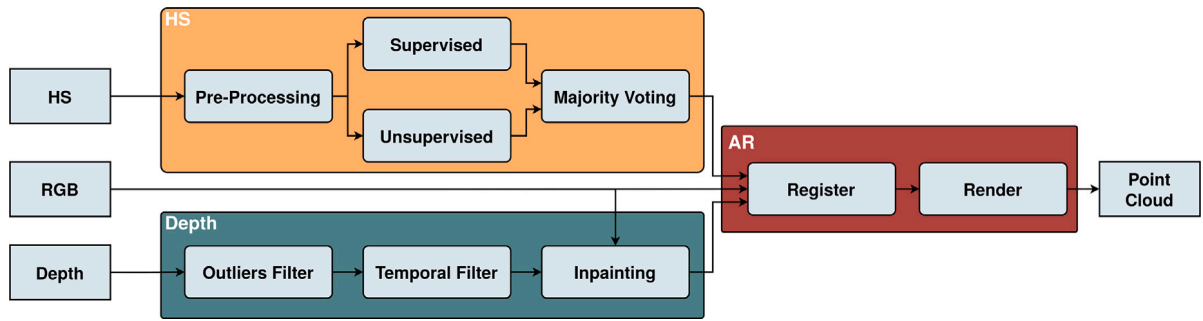

**Fig. 2.** System pipeline. The information sources are the HS information from the HS camera and the RGB and depth information from the LiDAR. The final result is a rendered video point cloud. The three main processing chains are: (in orange) HS processing, (in green) depth processing, and (in red) AR processing. Every box inside a processing chain refers to a process. (For interpretation of the references to color in this figure legend, the reader is referred to the web version of this article.)

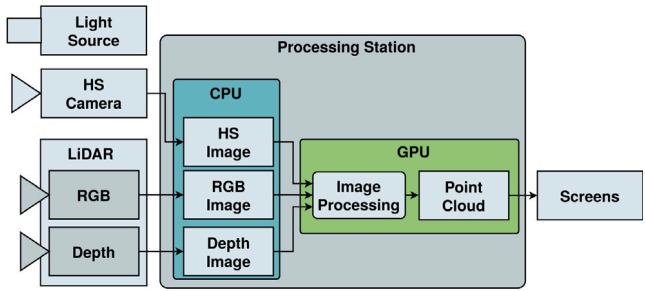

**Fig. 3.** Slimbrain physical system scheme and data flow.

### 3.2.1. Pre-processing

Using a HS snapshot camera requires the implementation of a pre-processing stage that transforms the raw captures into real HS cubes. The first step to do so is the conformation of the HS cube from the 2045 × 1085 mosaic raw capture to a 409 × 217 × 25 cube. However, in order to fit the GPU warp (a set of 32 contiguous threads in the GPU that constitutes the basic SIMD unit) size, the number of bands is set to 32, hence the size of the cubes is 409 × 217 × 32 (with 7 bands of padding). In addition, in order to make the best of the acceleration and proved that the quality is not compromised [39], the cube is stored using floating point numbers with 16 bits (FP16).

The previous process is encapsulated within the black-and-white calibration, where every pixel is independently remapped between the darkest and brightest values, given by a dark-and-white calibration capture. Once it is done, the calibrated values are then stored as a HS cube, using Band Interleaved Pixel (BIP) arrangement.

The calibrated cube is spectrally corrected using the correction matrix given by the manufacturer. This process requires a vector–matrix multiplication for every HS pixel, where the vector is a HS pixel with 32 components and the matrix is the so-called correction matrix. Then, every HS pixel is normalized by the mean squared value of all its bands, improving the classification accuracy in supervised learning. Finally, the HS image is transposed and unpadded to fit a band sequential order arrangement in memory. This operation does not change the data itself, but greatly improves the performance for the classification stage.

### 3.2.2. Classification

The classification is divided into three main stages: a supervised classification algorithm, an unsupervised clustering algorithm, and a majority voting final stage.

The supervised classification, based on a support vector machine (SVM) with a radial basis function kernel [41], processes every HS pixel with 25 bands in real-time video to produce a probability per class. In this work, every HS pixel is a sample with 25 features (the bands) that is independently classified to a class (in our examples, normally 4). To do so, a model is previously computed using pre-processed HS images captured in the surgical room and later labeled by expert neurosurgeons using a labeling tool [16].

The unsupervised classification, based on K-Means [42], generates up to 64 clusters from the pre-processed HS cube based on the pixel spectral similarity. The spectral similarity is computed by means of the L2 norm, a well-known metric that computes the euclidean distance between two spectral signatures (as they were vectors of $\lambda$ components). This algorithm starts initializing all the clusters to a single random pixel and then they grow iteratively given the metric until every pixel in the image is included in a cluster. In the end, the clusters only contain spectrally similar pixels.

Finally, the resulting classifications of both classifiers are mixed in a single classification map using a majority voting algorithm. For every cluster, a mean probability for each class is computed using the supervised classification probabilities. In this way, the noise generated by the per-pixel SVM approach is filtered and a smoother classification map is generated. To generate the final colored classification map, a color is assigned for each class (for instance, the tumor tissue class is red). Then, the color for each pixel is computed as a linear combination of the classes in that pixel, pondered by their probabilities (as shown in the results section, in Fig. 17).

## 3.3. Depth processing chain

The depth processing chain arises as a quality improvement stage for the depth information provided by the Intel Realsense L515 LiDAR. The use of such a device is intended for real-time applications, as other depth estimation techniques such as HS multiview systems require much higher processing times [43] and would greatly increase the size of the system. However, the LiDAR, based on the ToF technology [44], has some implicit problems that can be observed mainly in slanted borders or black areas. This happens when the reflected signal emitted by the transmitter cannot reach the receiver. An example of a LiDAR depth capture in the surgical room is presented in Fig. 4 along with its corresponding RGB capture. As observed, some black regions appear on the capture, meaning that the LiDAR has not captured information for that points, and hence the 3D model created will have holes or flying points. In this example and in general, for these captures, the errors occur in reflective objects and farther regions such as the ground or the bedsheets. In addition, some devices employed in the normal course of an operation emit signals that interfere with the depth capture. This results in the black line in the middle of the capture, which traverses the scene from left to right periodically.

The RGB image and the depth image are not captured from the same camera and hence they need to be aligned to be employed. Although the depth map captured by the LiDAR can be aligned with the RGB image, this process would entail even more artifacts in the final depth map given the different sizes of the images and their different point of view. For this reason, the RGB image is the one that is aligned to the position of the depth sensor. This fact has some implications in the depth processing chain (bottom part of Fig. 2). All its stages; outliers filter, temporal filter, and inpainting, works on the depth map captured by the LiDAR; while the RGB employed in the depth inpainting is the back-projection to the position of the depth sensor of the RGB captured. In addition, all this processing is performed using 16-bit precision with depths encoded as millimeters; the precision given by the manufacturer. This fact helps with the real-time acceleration of all the stages.

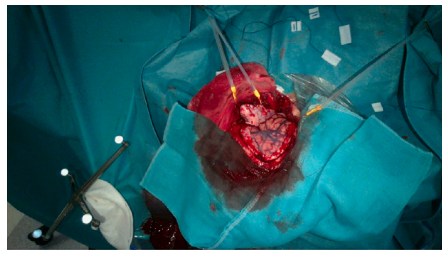

(a) Color image.

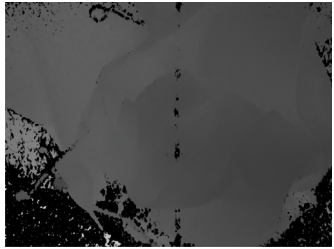

(b) Depth image.

**Fig. 4.** Capture by Intel Realsense L515 in a brain tumor operation. The depth map image represents the distance between the camera and the objects in the scene, where darker values refer to closer objects and whiter values to farther. This image shows areas where there is a lack of depth information in black pixels.

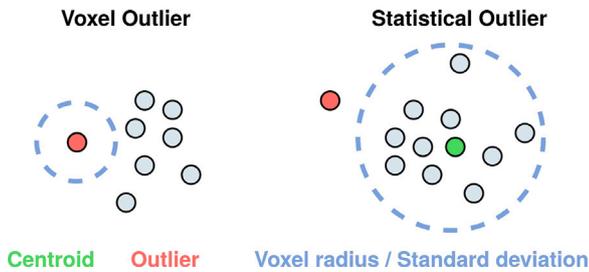

**Fig. 5.** Outlier filters employed in the depth processing chain. On the left, is the voxel outlier filter, where a 3D point is considered an outlier (in red) when there are no more points in a voxel radius (in blue). On the right, is the statistical outlier filter, where a 3D point is considered an outlier if it is far from the centroid (in green) more than $n_{std}$ standard deviations (in blue). Both voxel radius and $n_{std}$ are input parameters for the algorithms. (For interpretation of the references to color in this figure legend, the reader is referred to the web version of this article.)

*3.3.1. Outliers filter*

The outliers filter addresses the 3D points that are far from any other 3D point, as they are considered errors from the depth sensor and are displayed as flying points (see Fig. 4, depth image left bottom corner). This filter considers both statistical and voxel outliers, i.e., points that are far from an average centroid point more than $n_{std}$ standard deviations and points that are far from any other point in a radius of $r$ meters, respectively. These ideas are depicted in Fig. 5. They are performed through two GPU-accelerated functions that search in a $7 \times 7$ window per pixel in the depth map. These pixels are converted to 3D points given their depth and intrinsic matrix and then, the central pixel is evaluated to find whether it is an outlier or not.

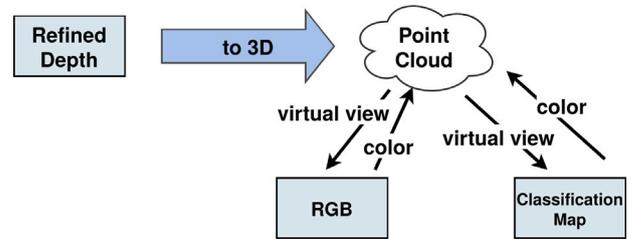

**Fig. 6.** Register process. Virtual view refers to the computer-generated image captured from a non-existent camera.

*3.3.2. Temporal filter*

The temporal filter tackles the task of reducing the inter-frame depth sensor noise while enhancing temporal coherence. Raw depth captures feature high fluctuations across the time for still objects. This is observed in the final 3D point cloud as waves that vary their high across the time several centimeters, around 5 cm. To remove them, the temporal filter averages the current frame depth with the previous frame depth only for close variations, under 5 cm. In this way, time fluctuations are greatly reduced and real movements are still captured.

*3.3.3. Depth inpainting*

This stage focus on inpainting the areas where the LiDAR cannot obtain information due to ToF inherent problems. These areas are displayed as black pixels in the depth map and holes in the 3D point cloud, worsening the subjective quality of the representation. As these areas are not the majority of the scene, the depth can be determined by an interpolation of the real information guided by the aligned RGB image. In this stage, the depth is treated as a grayscale image with noise that is filled by the surrounding grayscale values when there are similarities in the aligned RGB image.

*3.4. Registration and rendering*

The last step is the representation of all the processed information in a video point cloud. To do so, first, the information from the three different sources is registered on a single point cloud. Then, this point cloud is rendered using OpenGL [45], also providing the tools to navigate it (right part of Fig. 2).

*3.4.1. Depth, RGB and classification map registration*

The registration process revolves around the depth map image refined, which is captured from the point of view of the depth sensor. Using that information along with the intrinsic and extrinsic parameters of all the cameras involved, the registration is performed. The process is depicted in Fig. 6 and further explained below.

The first step is to convert the depth map into a point cloud in the world reference system by converting each 2D point with depth information into a 3D point [46]. This is performed using the following equation, where $P_w$ refers to a 3D point (capital p) in the world reference system (suffix $w$); K is the intrinsic matrix, R is the rotation matrix and t is the translation vector for the depth camera (suffix $depth$); $p(u, v, z)$ is a 2D point in the camera reference system and its depth.

$$P_w \begin{bmatrix} x \\ y \\ z \\ 1 \end{bmatrix} = \left( \frac{K_{depth} * [R_{depth}|t_{depth}]}{0\ 0\ 0\ 1} \right)^{-1} * p_{depth} \begin{bmatrix} u \\ v \\ z \\ 1 \end{bmatrix} \quad (1)$$

Then, the point cloud, without any color information, is colored using the RGB information from the RGB camera integrated into the LiDAR. To do so, the point cloud is projected to the RGB camera position using the same intrinsic parameters as the RGB camera, generating a virtual camera [46]. During that process, the corresponding index image for each point in the point cloud is generated, allowing to map the color from the RGB camera to the point cloud. This is expressed in the following formula, where $s$ refers to a scale constant, and the suffix $RGB$ refers to the RGB camera.

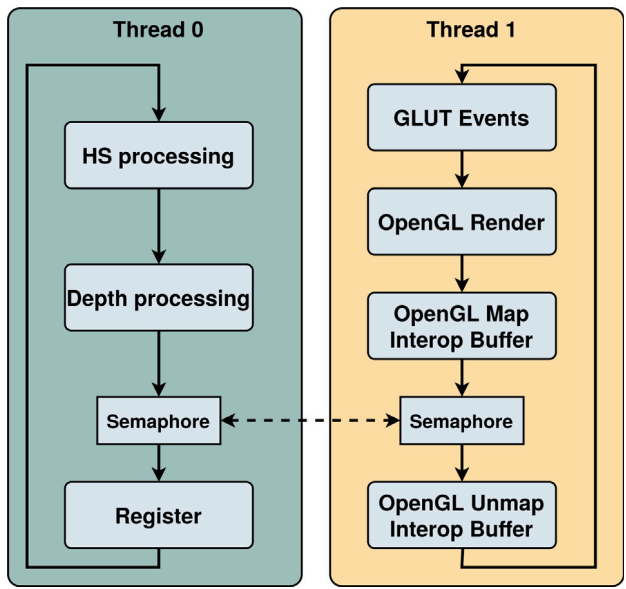

**Fig. 7.** Multithread processing-rendering.

$$\frac{1}{s} p_{RGB} \begin{bmatrix} u \\ v \\ s \\ 1 \end{bmatrix} = \left( \frac{K_{RGB} * [R_{RGB}|t_{RGB}]}{0 \ 0 \ 0 \ 1} \right) * P_w \begin{bmatrix} x \\ y \\ z \\ 1 \end{bmatrix} \quad (2)$$

Finally, the point cloud is colored with the classification map generated in the HS processing chain replicating the previous process with the correct intrinsic and extrinsic parameters. Given the HS camera's limited field of view (FOV), this process results in a point cloud mainly RGB, where only the region of interest, the exposed brain, has classification information.

#### 3.4.2. Rendering

The generated point cloud, already stored as a set of points with position and color information, is then represented using OpenGL and GLUT [47] in a 3D space. This is implemented as a multi-threaded software where the HS and depth processing run in one thread while the render is executed in another. This scheme allows the manipulation of the 3D view while the multi-modal captured video is being processed, with no appreciable latency.

To do so, the interoperability between OpenGL and CUDA [48] is exploited, i.e., a direct method to move information already stored in the GPU to an OpenGL buffer (that is also in the GPU). This allows saving the registering output directly on the OpenGL buffer, avoiding an extra copy. However, this copy needs to be synchronized with the OpenGL-GLUT render pipeline, forcing the use of a semaphore to ensure that the copy is only performed while the interoperability buffer exists. This idea is represented in Fig. 7.

### 4. Verification and results

The results section describes the tests performed on SLIMBRAIN to verify its usability in a surgical room during brain tumor resection operations. These tests comprise (i) physical and logistic system considerations, (ii) HS classification accuracy and time results, (iii) depth processing validation through quality and time results, and (iv) complete system verification in real operations.

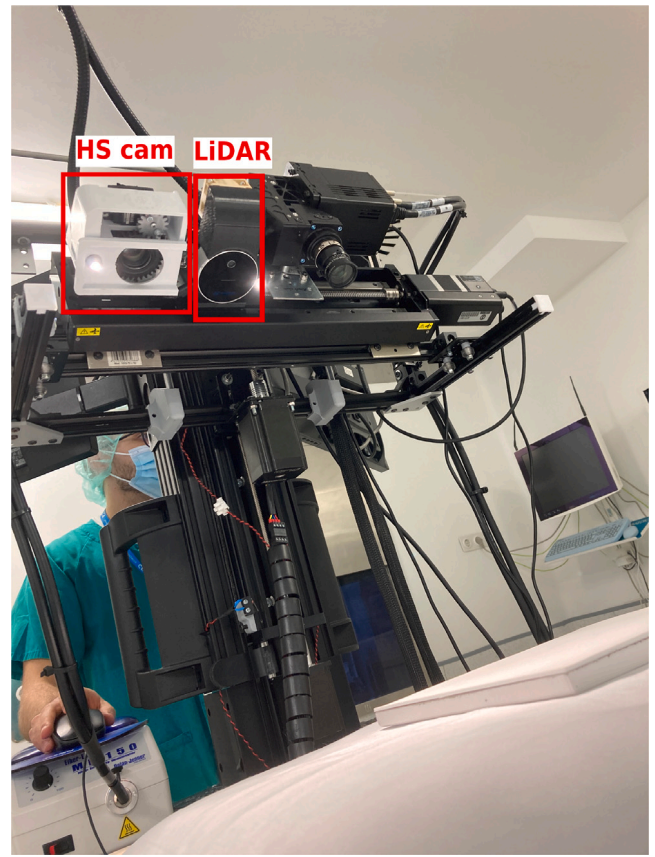

**Fig. 8.** Camera physical distribution. The image is captured in a real empty surgical room while experimenting with the white image calibration.

#### 4.1. Slimbrain physical system

This system has been constructed to be used in a real surgical room, where some constraints are present. The standalone system, including its main elements, described in Section 3.1, is mounted on a movable structure that allows an easy displacement within the surgical room. The cameras are mounted on a platen that can be moved and oriented toward the scene, allowing to employ the system in different operations and without any restriction on the region of interest. The reduced size and weight of the cameras, less than 0.5 kg, allow easy movement along x–y axes. This is depicted in Fig. 8.

#### 4.2. HS processing verification

In order to test the HS processing, 5 video sequences captured during real brain tumor operations of different patients are employed. In all cases, the patients suffered glioblastoma multiforme (GBM) grade IV with non-mutated isocitrate dehydrogenase (IDH). The first frame of these videos is labeled by neurosurgeons to create the ground truth needed for model generation and validation. After, using the ground truth of these 5 video sequences, a model is trained using 80% of the HS pixels from all the patients. The model accuracy is assessed using the rest 20% pixels not selected in the training phase. In this way, both stages share patients but employ different pixels. The model shown in Fig. 9 yields an area under the curve (AUC) [49] of 95.27% globally and 95.17% for the class tumor. These numbers show a good classification on average for the tested pixels, however, subjectively, for some images the classification is not successfully achieved.

Although the objective results are extracted from the one-class representation (a pixel has a class label, the one with more probability),

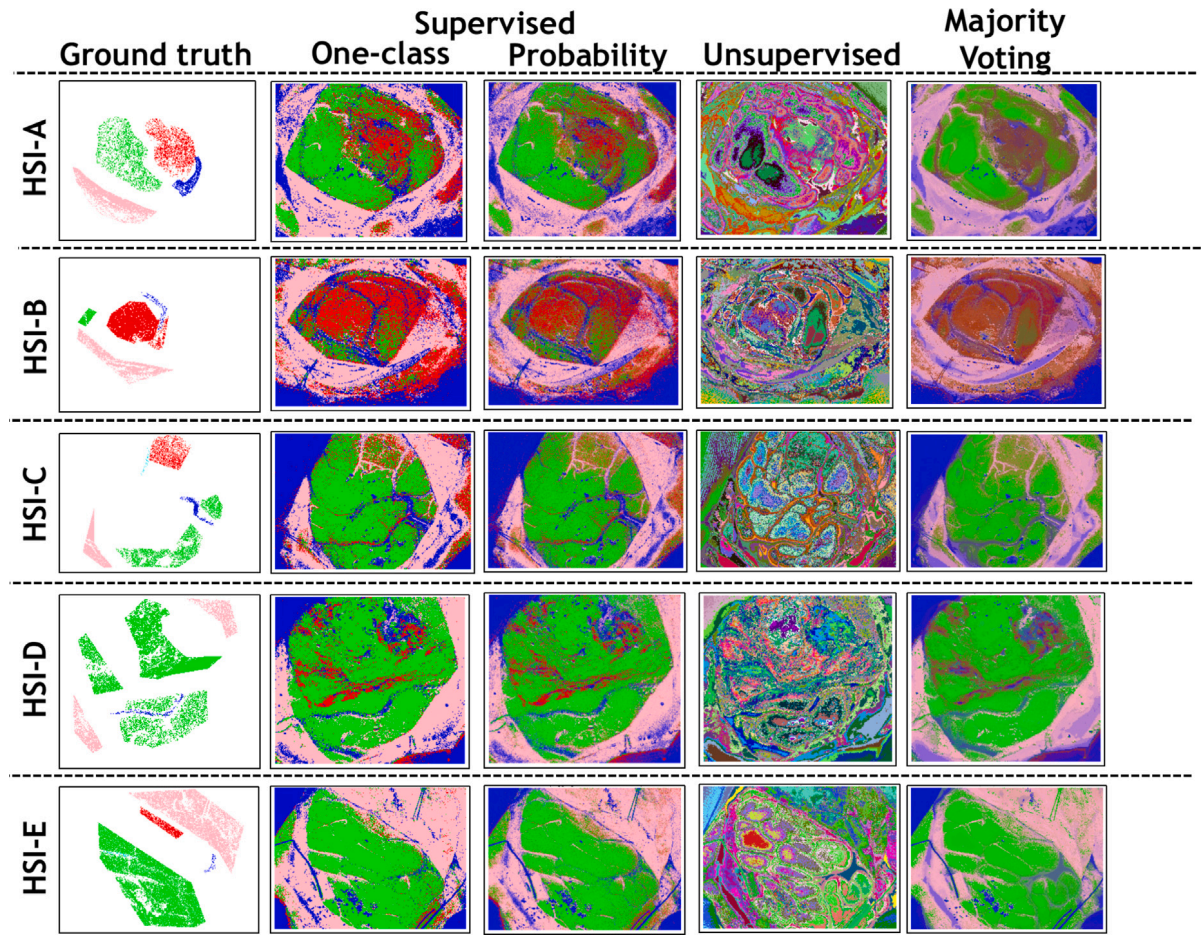

**Fig. 9.** Classification results. Red pixels refer to tumor tissue, green pixels to healthy tissue, blue pixels to blood, and pink pixels to dura-mater. Each row represents a different brain tumor capture, while each column depicts a different type of information. From left to right: the ground truth labeled by neurosurgeons, the one-class and probability classification maps result of the supervised training, the cluster image produced by the unsupervised algorithm with a different color per cluster, and the result of the majority voting, mixing supervised and unsupervised classification maps. (For interpretation of the references to color in this figure legend, the reader is referred to the web version of this article.)

the images shown to the neurosurgeons are the ones at the end of the HS processing chain. All the steps in this chain are represented in Fig. 9, where the majority voting column is the one presented to the neurosurgeons. This is intended for giving a better representation taking into account that the supervised chain produces a probability and not only a higher-probability class. These probabilities contain useful information that can improve the visualization. This probability image is mixed with the result of the unsupervised chain, which finds spectral clusters across the image. As depicted in Fig. 9, the final results vary from the one-class classification, enhancing the visualization through a smoother representation. In these images, not pure colors are represented, they are the combination of four basic colors, each one assigned to a class: red for tumor, green for sane, blue for blood, and pink for dura-mater. This combination depends on the probability of the supervised chain in the probability map and in the average of the cluster in the majority voting classification map.

These results show different aspects of the processing chain. Focusing on the one-class supervised learning results, almost all the captures present a good global correlation between the ground truth and the prediction, which explains the good AUC results. However, it is important to remark the ground truth sparsity, which tends to increase the objective metrics while reducing the subjective quality in the rest of the image. This effect can be mainly observed in some of the healthy areas where the classifier predicts tumor tissue, although the rest of the classes are correct. For example, in HSI-A, HSI-B, and HSI-D, the classifier tends to incorrectly predict tumor tissue in healthy areas out of the ground truth region.

The misclassification effect between healthy and tumor tissue is reduced when combining the unsupervised results with the supervised ones. Focusing on HSI-A, the real tumor region becomes clearer while the false region turns greener. In the other images, this stage is not able to improve the tumor-supervised learning results, even eliminating true healthy tissue in some cases such as HSI-B, in the lower region. However, regions like veins or arteries and their frontier with other types of tissue are better defined. This can be seen for example in HSI-A or HSI-E, where some blood vessels appear even when they were completely hidden in the supervised training result. For the rest of the images, a similar result can be observed.

All the previous results are contingent on their implementation in real-time video, as it is required for the complete AR system. For this reason, it is important to measure the time execution of these stages to ensure a correct visualization. The processing time measured is the result of an average of 200 executions of every stage in the GPU (the specific model is in Section 3.1), while the complete system is running.

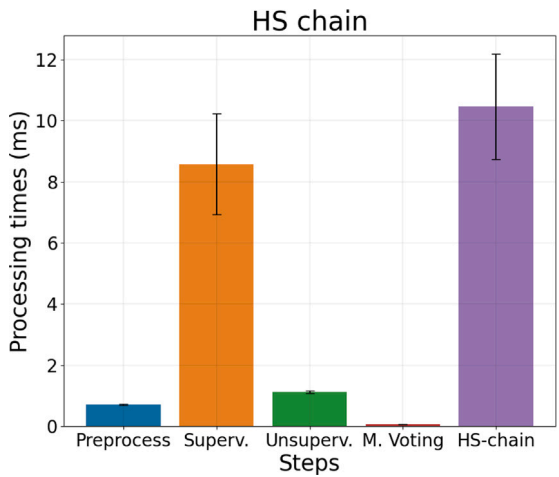

**Fig. 10.** HS chain per-frame-time results. The vertical bar means average whilst the black line in each bar refers to the standard deviation in the measurement.

This is an important factor, as OpenGL rendering also uses the GPU and hence, some performance loss is expected. In addition, it is important to remark that this specific combination of HS camera and light source results in a need of an exposure time of 70 ms, setting the maximum attainable FPS to around 14 FPS. Time results are included in Fig. 10.

As can be seen, processing times in this chain are much lower than these 14 FPS, consuming around 11 ms, i.e., near 90 FPS. This enables its use in the AR within the system.

*4.3. Depth processing verification*

The depth processing verification aims to justify the use of a single LiDAR to extract the depth information and the need of a processing chain that improves the raw LiDAR depth maps. In order to do this, 5 static sequences were captured in a laboratory using the same LiDAR mounted in an xyz stage (movement up to $0.8 \times 1.2 \times 0.85$ meters with an accuracy of less than 1 mm, 6400 motor steps per millimeter for each axis). The experiments for every sequence, consist of capturing a $5 \times 5$ square array from positions separated 1 cm and with a distance to the farthest and nearest plane of 1 m and 0.3 m, respectively (as in the operating room). The capture includes RGB and depth images not aligned, i.e., both acquired from their corresponding sensor position. With that captures, the quality of the depth maps is validated through an assessment chain based on (i) MPEG reference video synthesis (RVS) [50], a SoTA view synthesizer able to generate virtual views in any position given any number of RGBD captures and (ii) IV-PSNR [51], an assessment tool specialized in the evaluation of syntheses. The LiDAR captures are also compared to a SoTA depth estimation tool, MPEG-I Depth Estimation Reference Software (DERS) [52].

The assessment chain has three paths: (i) DERS, (ii) LiDAR, and (iii) corrected LiDAR. These paths refer to the different solutions evaluated to solve the problem of generating an AR for this system. They are depicted in Figs. 11, 12 and 13, respectively. The three modalities are detailed in the following paragraphs.

In the first modality (Fig. 11), MPEG-I DERS is employed as a high-quality depth estimation tool present in the SoTA. The use of this software is intended for evaluating whether a LiDAR is a good candidate compared to a classic RGB multiview depth estimation algorithm or not. To this end, the system is mimicked replacing the LiDAR with an RGB multiview camera array with 5 cameras (a cross-shape array centered on the central view) where DERS is employed to generate a depth map in the position of the central view. This depth map, estimated in the RGB camera position has the benefit of having a $1920 \times 1080$ spatial resolution, as this is the RGB size image. This depth estimation along with the RGB central view of the array constitutes the central RGBD image (the yellow box in Fig. 11). With this image, an RVS virtual view is generated in the rest of the positions. Finally, the syntheses are compared to the real RGB images using IV-PSNR, obtaining an IV-PSNR value for the DERS modality.

The second modality (Fig. 12) employs directly the depth map produced by the LiDAR, with the inconvenience that the depth map images, captured by the depth sensor, have a $1024 \times 768$ resolution. Apart from the lower resolution, the depth is captured from a point of view different from the RGB images. For this reason, the assessment chain slightly varies from the previous one. Mimicking the process in the real system, the captured images are projected to the position of the depth sensor, obtaining an array of RGB projected images. The central view of that array joint to its LiDAR depth constitutes the RGBD image under test (the blue box in Fig. 12). Using that image, RVS is employed to synthesize an image in all the other real positions, which already have a projected RGB version of them. Finally, the projected RGB and the synthesized RGB are compared for every view. This comparison is performed only in the areas where the image exists; some parts of the image are black due to the projection process, and they lack information. These parts are different in the synthesis and the projected RGB so only the overlapping pixels are evaluated. In this way, IV-PSNR gives a result considering only the real visible parts, although the FoV is different. This process ends up with an average IV-PSNR value for the LiDAR modality.

The third modality (Fig. 13) makes use of the corrected depth to perform exactly the same process as in the previous modality (the red box in Fig. 13 is the corrected depth). The intention of this modality is comparing the raw depth produced by the LiDAR to its corrected counterpart to see if a correction in the depth is justified.

This assessment chain allows a fair comparison between the syntheses generated by the three different depth maps. The results are presented in Fig. 14 in a box plot. This chart compares the IV-PSNR results for four different sequences and the three depth modalities in each one. For every sequence and modality, all the side views (24 views) are considered within the box plot, showing their average, standard deviation, maximum and minimum.

In the first place, results show that on average DERS obtains around 32.5 dB, 1.7 dB less than both LiDAR versions, although its maximum sometimes reaches the same level as IV-PSNR. Also, the standard deviation of this modality is much more pronounced, finding these differences in cameras near or far from the central view. This result shows that the difference between the maximum and minimum IV-PSNR in DERS is almost 6 dB, meaning that for farther syntheses, the image quality heavily drops. Despite the difference in IV-PSNR in favor of the LiDAR, it is worth remarking that the syntheses produced by DERS feature a spatial resolution of $1920 \times 1080$, whilst the LiDAR depth is only $1024 \times 768$, more than 2.5 times less. In order to fairly compare the objective quality using the IV-PSNR tool, this calculation has been performed resizing DERS syntheses to $1024 \times 768$.

In the second place, the correction stage for the depth information seems unclear given the results. In most of the sequences (excluding seq2), the average, standard deviation, maximum and minimum present a slight increase of less than 1 dB. This happens for every view, to a greater or lesser extent.

Given the slight quality increase in terms of IV-PSNR, subjective results are attached in Fig. 17 in one of the worst IV-PSNR results sequence-view, and in Fig. 18 in one of the best IV-PSNR results sequence-view. In both figures, the differences between the three modalities are subtle and hard to differentiate. Comparing DERS with L515, differences in borders can be appreciated, for example in the edge of the white plastic skull (in both figures), which seems more blurred and larger than the ground truth. However, from the subjective point of view, DERS gives a better representation than L515 as it lacks small holes and overall artifacts. The processed depth map from L515 gets rid of such holes and artifacts, as clearly seen in the plastic skull (in both figures), the black pyramid (in Fig. 17, the top edge of the blue plastic brain (in Fig. 18), or the gray object on the left (in Fig. 18). From the authors' point of view, DERS and L515-corrected have similar subjective results, without any significant artifacts.

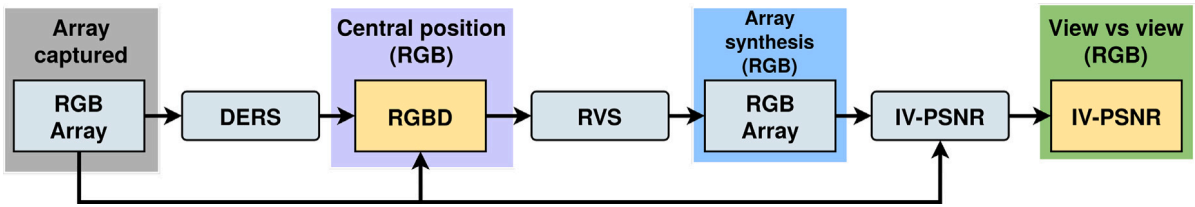

**Fig. 11.** Depth assessment chain for DERS modality. Only using the RGB array captured, a depth in the central position is generated. This is employed to synthesize a view in all the other positions, where they are evaluated using IV-PSNR. (For interpretation of the references to color in this figure legend, the reader is referred to the web version of this article.)

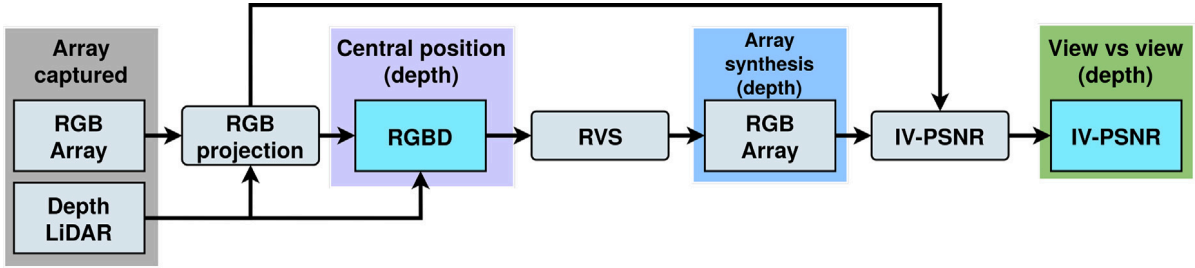

**Fig. 12.** Depth assessment chain for LiDAR modality. Using the LiDAR depth and projected RGB in the central position, a synthetic view is generated in all the other positions. They are compared to the rest of projected RGB views using IV-PSNR. (For interpretation of the references to color in this figure legend, the reader is referred to the web version of this article.)

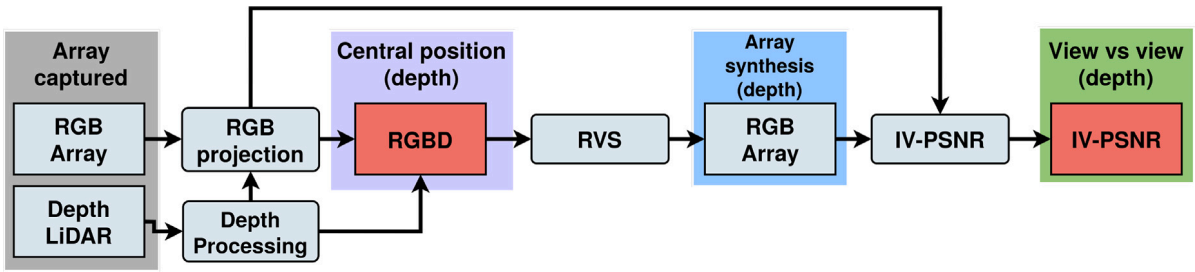

**Fig. 13.** Depth assessment chain for corrected LiDAR modality. Using the corrected LiDAR depth and projected RGB in the central position, a synthetic view is generated in all the other positions. They are compared to the rest of projected RGB views using IV-PSNR. (For interpretation of the references to color in this figure legend, the reader is referred to the web version of this article.)

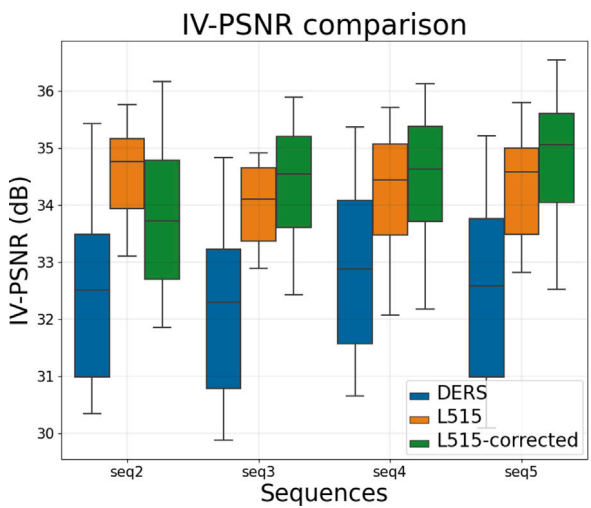

**Fig. 14.** IV-PSNR results for the three modalities.

**Table 1**
Processing time for the three depth modalities.

| Modality | Average time (s/frame) |
| --- | --- |
| DERS | 700 |
| LiDAR | 0.033 |
| LiDAR-corrected | 0.033 + 0.003 |

The last factor to consider is the processing times for each modality. Table 1 presents these results and Fig. 15 depicts the processing time for the depth chain. This table shows that the processing time for DERS is clearly far from being employed in a real-time video application, whilst the raw LiDAR offers an acceptable solution at 30 FPS. In the case of the corrected LiDAR, the processing time added to the acquisition is around 2.9 ms, lowering the output frame rate to 27.8 FPS. As depicted in Fig. 15, the majority of this time corresponds to the outlier filtering.

Overall, results obtained show that the LiDAR corrected version offers subjective quality results near a high-quality depth estimation software at around 28 FPS. Although this justifies its use in the AR system, the spatial resolution obtained is significantly lower than the one obtained for the depth estimation software.

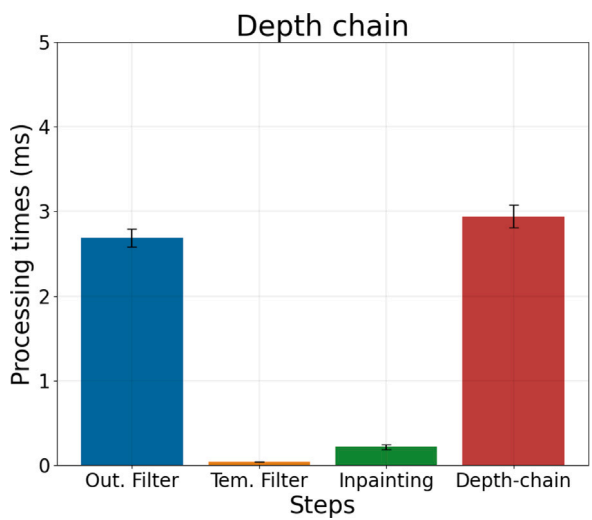

**Fig. 15.** Depth chain per-frame-time results. The vertical bar means average whilst the black line in each bar refers to the standard deviation in the measurement.

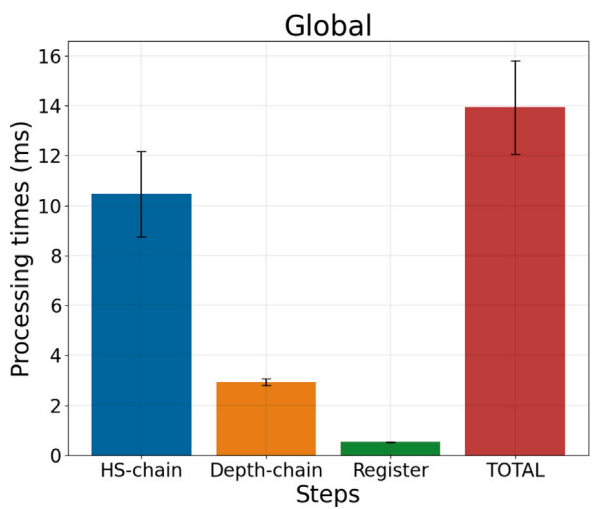

**Fig. 16.** Global chain per-frame-time results. The vertical bar means average whilst the black line in each bar refers to the standard deviation in the measurement.

*4.4. Real environment verification*

In this final section, the whole system is tested in a surgical room during real brain tumor resection operations. Given the inner restrictions of a real operation, few tests can be performed under its course. The ones included in this work are the end-to-end frame rate attained and the real-time video point clouds generated during the operations. The time results are included in Fig. 16, where the three processing chains are summarized.

The final point clouds are included in Fig. 19 as well as the RGB image and the point cloud without correction from the same point of view. They show only a point of view of the point cloud, although free navigation is possible.

These images show the final result presented to the neurosurgeons, which can navigate the model in real-time at the same time the information is recorded. Time results show a maximum attainable frame rate of around 21 FPS (33 ms LiDAR capture + 14 ms processing), although the HS camera requirements limit this capture rate to 14 FPS due to the exposure time needed in the surgical room. In addition, the quality results presented show a tool able to overlay the classification information produced by HS cameras on an RGB point cloud during a real tumor resection operation. The point cloud generated is also corrected to obtain a better representation, something that can be seen in Fig. 19 for example in the black line that traverses HSI-B or the black artifacts in HSI-D in the dura-mater (pink) right area.

Although the system has been technically validated in several real operations, a thorough clinical validation remains to be done. In this work, the HS classification has been verified through the comparison between the classification maps and the ground truth labeled by the neurosurgeons. However, the complete AR system opens the door to a more precise clinical evaluation thanks to the better correspondence between the classification map and the histopathological tests of the resected tissue. This work will be addressed in the near future through methodological clinical tests led by the neurosurgery service, where a comparison will be made between the resected tissue and the classification map results.

## 5. Conclusions

This paper presents SLIMBRAIN, a real-time AR system that processes and displays brain tumor HS classification results. This system aims to enlarge the current state-of-the-art where HS images are employed to differentiate tissues in the exposed surface of the brain. This technology is able to help neurosurgeons to determine where is the tumor during the course of the operation, reducing the amount of healthy tissue that is resected and hence improving patients' subsequent quality of life. However, this technology presents two main drawbacks: (i) employing high-resolution linescan cameras entails large capture times and prevents its continuous utilization during the surgery course, and (ii) the processed classification map generated from the HS camera presents an image difficult to correlate with the real brain surface, reducing the effectiveness of the technology in borders and confusing regions.

SLIMBRAIN addresses these problems by employing a HS snapshot camera plus a LiDAR device to generate an AR visualization where the information is captured and processed in real-time. By means of these real-time video cameras and GPU accelerators, the fully accelerated processing chain is able to attain up to 21 FPS while the AR is presented to the neurosurgeons. Objective classification results employing an SVM classifier show a global and tumor AUC of around 95% although the subjective results present deficiencies. The granularity is solved by a K-Means filtering that smooths similar spectral regions and enhances the visualization. However, for some images, tumor tissue is considerably confused with healthy tissue, highly degrading the output quality. Regarding the depth quality, results show that a LiDAR plus a slight correction can produce an AR with similar subjective results as a high-quality depth estimation algorithm in video real-time. This allows a successful registration between the classification and the RGBD image, integrating the classification results over a wider RGB capture that improves the neurosurgeons' scene understanding. This representation also features scene navigation, enhancing the region of interest visualization.

Future work will address the improvement of the classification chain to reduce deficiencies in tumor regions. The database will be gradually enlarged with new patients and captures while new techniques and algorithms employing the scarce information from snapshot cameras will be developed. In addition, the system will be tested in more real operations performing a clinical evaluation through a subjective user evaluation and histopathological tests to assess the feasibility of the system in the health-care domain.

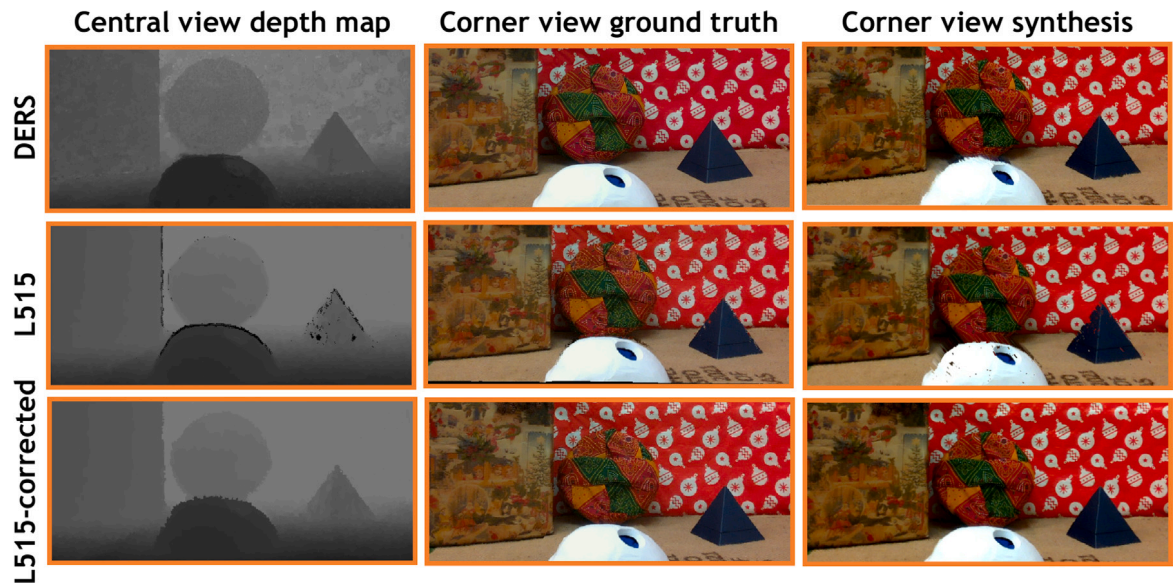

**Fig. 17.** Subjective results in synthesis for one of the worst IV-PSNR view. Detail from seq2 left-up corner view. (For interpretation of the references to color in this figure legend, the reader is referred to the web version of this article.)

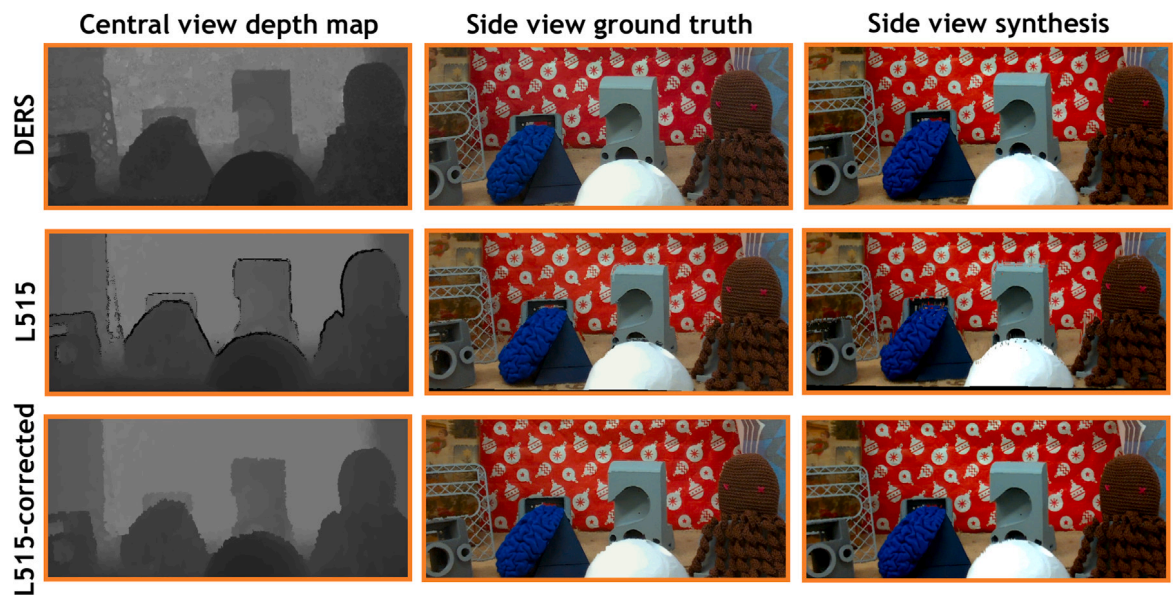

**Fig. 18.** Subjective results in synthesis for one of the best IV-PSNR view. Detail from seq5 next-to-central view. (For interpretation of the references to color in this figure legend, the reader is referred to the web version of this article.)

**Declaration of competing interest**

The authors declare that they have no known competing financial interests or personal relationships that could have appeared to influence the work reported in this paper.

**Data availability**

Data will be made available on request.

**Acknowledgments**

The authors would like to thank to all the staff of the neurosurgery operating room of the Hospital Universitario 12 de Octubre in Madrid for their help and support during the image capture process.

**Funding**

This work was supported by both the Regional Government of Madrid through NEMESIS-3D-CM project (Y2018/BIO-4826) and by the Spanish Ministry of Science and Innovation through TALENT project (PID2020-116417RB-C41).

**Institutional review board statement**

The study was conducted according to the guidelines of the Declaration of Helsinki, and approved by the Research Ethics Committee of Hospital Universitario 12 de Octubre, Madrid, Spain (protocol code 19/158, 28 May 2019).

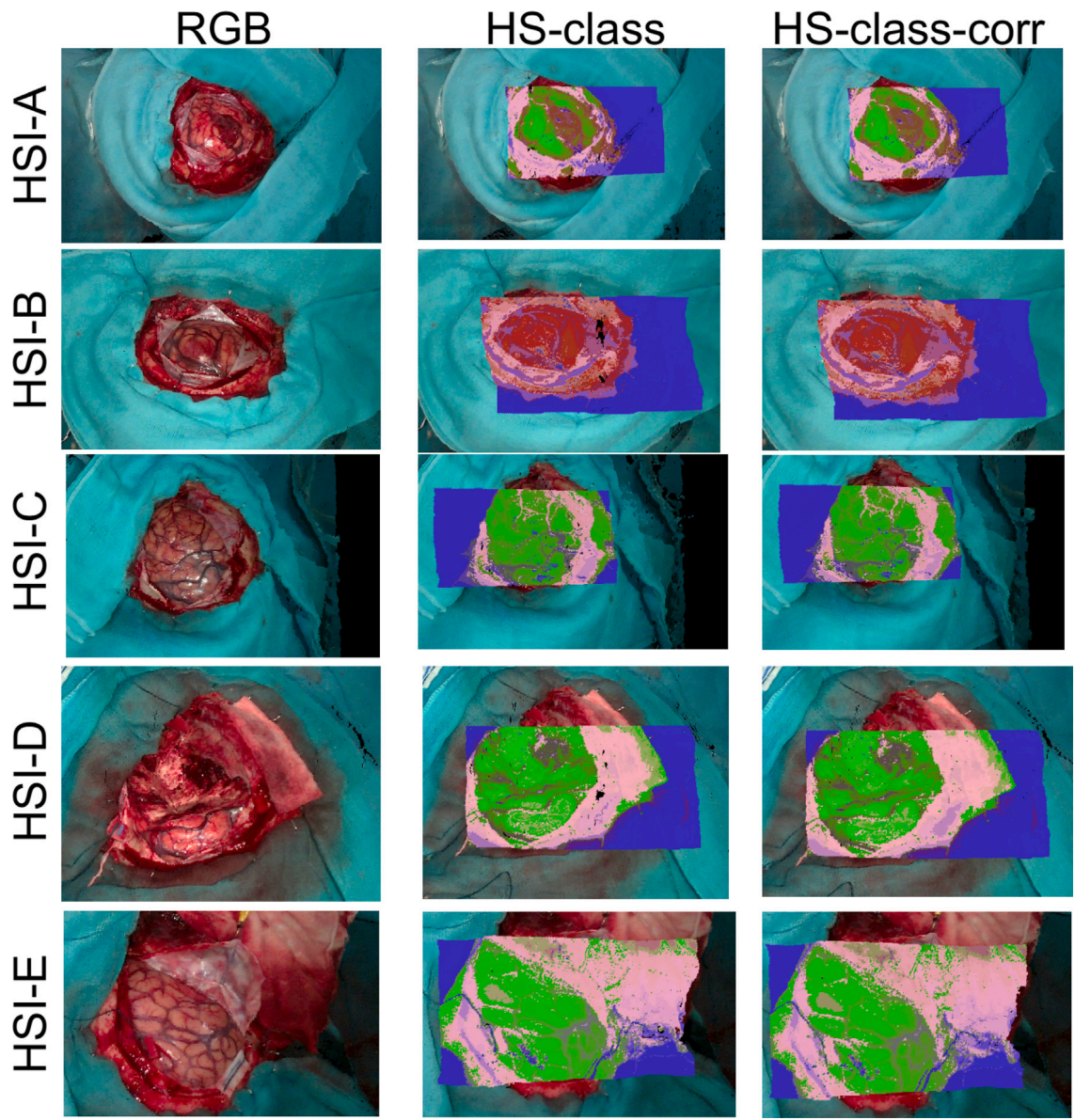

**Fig. 19.** Point cloud results in real operations. Each row refers to a different tumor operation while each column refers to a type of visualization. From left to right: RGB point cloud, RGB point cloud with HS classification information, and RGB point cloud with HS classification information and depth correction. (For interpretation of the references to color in this figure legend, the reader is referred to the web version of this article.)

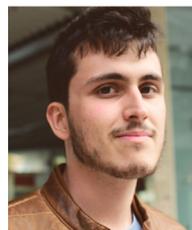

**Jaime Sancho** received his M.Sc. degree in Systems and Services Engineering for the Information Society from Universidad Politécnica de Madrid (UPM), Spain, in 2018. He is currently a PhD student at the Electronic and Microelectronic Design Group (GDEM) in the Software Technologies and Multimedia Systems for Sustainability (CITSEM) Research Center, UPM. His research interests include biomedical real-time systems and immersive computer vision technologies.

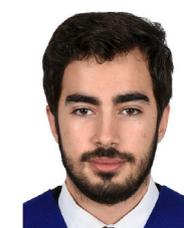

**Manuel Villa** received his master degree in Internet of Things (IoT) and bachelor degree in telecommunications from Universidad Politécnica de Madrid (UPM), Spain, in 2021 and 2020 respectively. He is currently a PhD student at the Electronic and Microelectronic Design Group (GDEM) in the Software Technologies and Multimedia Systems for Sustainability (CITSEM) Research Center, UPM. His research interests include biomedical real-time systems and immersive computer vision technologies.

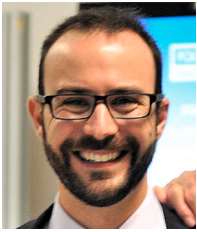

**Miguel Chavarrías** received the B.S. and M.S. degrees in telecommunications engineering and the Ph.D. degree from the Universidad Politécnica de Madrid (UPM), Spain, in 2011, 2012, and 2017, respectively. Currently, he is Associate Professor of UPM. He has been with the Electronics and microelectronics Design Group (GDEM) since 2011, and with the Software Technologies and Multimedia Systems for Sustainability (CITSEM) Research Center, since its creation in 2012. He has authored more than 30 journal articles and conference papers. He has participated in more than 15 Research and Development industry projects. He was a guest researcher at Institut d'Electronique et des Technologies du numérique (IETR) laboratory of the Institut National des Sciences Apliquées (INSA) in Rennes (France) during the springs of 2012 and 2015. He received the extraordinary doctorate award granted by the academic programme and IEEE Consumer Electronics Society Chester Sall Award for the third place best paper in the IEEE Transactions on Consumer Electronics, in 2014. His research interests include edge-computing, hyperspectral imaging and machine learning.

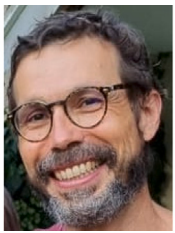

**Dr. Eduardo Juárez** received his PhD degree from the École Polythecnique Fédéral de Lausanne (EPFL) in 2003. From 1994 to 1997, he worked as researcher at the Digital Architecture Group of UPM and was a visiting researcher at the ENST, Brest (France) and the University of Pennsylvania, Philadelphia (USA). From 1998 to 2000, he worked as Assistant at the Integrated Systems Laboratory (LSI) of the EPFL. From 2000 to 2003, he worked as Senior Systems Engineer at the Design Centre of Transwitch Corp. in Switzerland, while continuing his research towards the PhD at the EPFL. In December 2004, he joined the Universidad Politecnica de Madrid (UPM) as a postdoc. Since 2007, he is Associate Professor at UPM. His research activity is mainly focused on (1) hyperspectral imaging for health applications, (2) real-time depth estimation and refinement and (3) heterogeneous high performance computing. He is co-author of one book and author or co-author of more than 100 papers and contributions to technical conferences. He has participated in more than 15 competitive research projects and 20 non-competitive industrial projects.

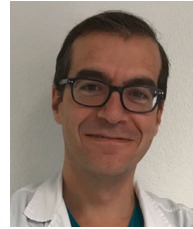

**Prof. Alfonso Lagares** is Professor in Neurosurgery in Universidad Complutense de Madrid and Head of Department of Neurosurgery in Hospital 12 de Octubre. He received his PhD in Neuroscience in 2004 in Universdad Autonoma de Madrid achieving Doctorate Extraordinary Prize. He is the coordinator of a research group in the research Institute imas12. His lines of investigation includes: (1) Prognostic models in traumatic brain injury and subarachnoid hemorrhage; (2) New tools for prognostication in different neurosurgical pathologies including radiological tools such as volumetric CT, conventional MR as well as diffusion and tractography MR for assessing white matter integrity in diffuse axonal injury; (3) Discovery of new biomarkers in the diagnosis and prognosis of head injury; (4) Design of new tools for improving tumor resection. He is author or co-author of more than 250 papers, received more than 4700 citations, and has participated in 20 competitive projects.

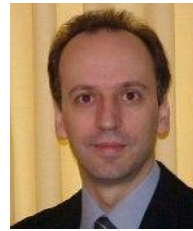

**Prof. César Sanz** received the Ph.D. degree in Telecommunication Engineering in 1998, from the Universidad Politécnica de Madrid (UPM). Since 1985, he has been a faculty member at the UPM, where he is currently a Full Professor. He has been the Director of the ETSIS de Telecomunicación, UPM, from 2008 to 2017. He leads the Electronic and Microelectronic Design Group (GDEM) since 1996, involved in R&D projects with private companies and public institutions. He is author and/or co-author of 10 books, an international patent and more than 100 papers and contributions to technical conferences. He has participated in more than 80 R&D projects. Since 2021, he is the director of CITSEM research center at UPM. His research interests include electronic design applied to video coding and hyperspectral imaging.